\documentstyle[12pt,epsf]{article}
\newcommand{\be}{\begin{equation}}
\newcommand{\ee}{\end{equation}}

\newcommand{\beqa}{\begin{eqnarray}}
\newcommand{\eeqa}{\end{eqnarray}}
\newcommand{\nn}{\nonumber}

\newcommand{\eqref}[1]{(\ref{#1})}

\newcommand{\one}{1\!\!1}

\def\boxit#1{\vbox{\hrule\hbox{\vrule\kern8pt
\vbox{\hbox{\kern8pt}\hbox{\vbox{#1}}\hbox{\kern8pt}}
\kern8pt\vrule}\hrule}}
\def\mathboxit#1{\vbox{\hrule\hbox{\vrule\kern8pt\vbox{\kern8pt
\hbox{$\displaystyle #1$}\kern8pt}\kern8pt\vrule}\hrule}}

\def\IB{\relax\hbox{$\inbar\kern-.3em{\rm B}$}}
\def\IC{\relax\hbox{$\inbar\kern-.3em{\rm C}$}}
\def\ID{\relax\hbox{$\inbar\kern-.3em{\rm D}$}}
\def\IE{\relax\hbox{$\inbar\kern-.3em{\rm E}$}}
\def\IF{\relax\hbox{$\inbar\kern-.3em{\rm F}$}}
\def\IG{\relax\hbox{$\inbar\kern-.3em{\rm G}$}}
\def\IGa{\relax\hbox{${\rm I}\kern-.18em\Gamma$}}
\def\IH{\relax{\rm I\kern-.18em H}}
\def\IK{\relax{\rm I\kern-.18em K}}
\def\IL{\relax{\rm I\kern-.18em L}}
\def\IP{\relax{\rm I\kern-.18em P}}
\def\IR{\relax{\rm I\kern-.18em R}}
\def\IZ{\relax\ifmmode\mathchoice
{\hbox{\cmss Z\kern-.4em Z}}{\hbox{\cmss Z\kern-.4em Z}}
{\lower.9pt\hbox{\cmsss Z\kern-.4em Z}} {\lower1.2pt\hbox{\cmsss
Z\kern-.4em Z}}\else{\cmss Z\kern-.4em Z}\fi}

\def\II{\relax{\rm I\kern-.18em I}}

\def\CA {{\cal A}}

\def\CL {{\cal L}}

\begin{document}

\setlength{\baselineskip}{7mm}
\begin{titlepage}
\begin{flushright}
{\tt NRCPS-HE-01-08} \\
January, 2008
\end{flushright}

\vspace{1cm}

\begin{center}
{\Large {\it   Duality Transformation  \\
of\\
non-Abelian Tensor Gauge Fields }
}
\vspace{1cm}

{ { Sebastian Guttenberg }\footnote{guttenb(AT)inp.demokritos.gr }}
and
{ {George  Savvidy  }\footnote{savvidy(AT)inp.demokritos.gr}}

\vspace{1cm}

\item {\it Institute of Nuclear Physics,} \\
{\it Demokritos National Research Center }\\
{\it Agia Paraskevi, GR-15310 Athens, Greece}

\end{center}

\vspace{1cm}

\begin{abstract}
For non-Abelian tensor gauge fields we have found an alternative form of
duality transformation, which has the property that the direct and the
inverse transformations coincide. This duality transformation
has the desired property that the direct and the inverse transformations
map Lagrangian forms into each other.

\end{abstract}

\end{titlepage}

\pagestyle{plain}

\section{\it Introduction }

In the recent decades string field theories and higher-spin field theories
became a subject of intensive research. One of the purposes of this
development is to find out an effective method for calculating {\it off-shell}
scattering amplitudes of high spin fields.

In the field theoretical approach the Lagrangian and S-matrix formulations of {\it free }
massless Abelian tensor gauge fields have been  constructed in
 \cite{fierz,fierzpauli,yukawa1,wigner,schwinger,
Weinberg:1964cn,chang,singh,fronsdal,deWit:1979pe,Curtright:1980yk}.
The  {\it interaction } of higher-spin fields
has been studied  in the light-cone formalism and
in the covariant formulation of the theories
\cite{Bengtsson:1983pd,Bengtsson:1983pg,Bengtsson:1986kh,berends,
Berends:1984wp,Metsaev:2005ar,Metsaev:2007rn,Boulanger:2006gr,Francia:2007qt}.
In string field theory
the interaction of higher-spin fields has been studied in
\cite{Thorn:1985fa,Siegel:1985tw,Witten:1985cc,
Siegel:1988yz,Arefeva:1989cp,Taylor:2003gn,
Taylor:2006ye,Berkovits:2003ny,Schnabl:2005gv}.
The  interacting field theories in anti-de Sitter space-time background
are reviewed in \cite{Bekaert:2005vh,Engquist:2002vr,Sezgin:2001zs}.

In general the concept of local gauge
invariance allows one to define the non-Abelian gauge fields \cite{yang},
to derive their  dynamical field equations
and to develop a universal point of view on matter interactions
as resulting from the exchange of spin-one gauge quanta. A possible
extension of the gauge principle which defines the
interaction of high-spin gauge fields has been made recently in \cite{Savvidy:2005fi}.
The resulting gauge invariant Lagrangian
defines cubic and quartic self-interactions of charged gauge quanta
carrying a spin larger than one
\cite{Savvidy:2005fi,Savvidy:2005zm,Savvidy:2005ki}.

Recall that in these publications it was found that there exists
not one but  a   pair, $\delta$ and $\tilde{\delta}$,
of complementary  non-Abelian gauge transformations acting on
tensor gauge fields  of the  rank s+1:
$
A^{a}_{\mu\lambda_1 ... \lambda_{s}}
$.
These are totally symmetric tensors with respect to the
indices $  \lambda_1 ... \lambda_{s}  $, but {\it a priori}
have no symmetries with respect to the first index  $\mu$.
The {\it extended gauge transformation} $\delta_{\xi}$
has the following form \cite{Savvidy:2005fi,Savvidy:2005zm,Savvidy:2005ki}:
\beqa\label{polygaugefull}
\delta_{\xi} A^{a}_{\mu} &=& ( \delta^{ab}\partial_{\mu}
+g f^{acb}A^{c}_{\mu})\xi^b ,~~~~~\\
\delta_{\xi} A^{a}_{\mu\lambda_1} &=&  ( \delta^{ab}\partial_{\mu}
+  g f^{acb}A^{c}_{\mu})\xi^{b}_{\lambda_1} + g f^{acb}A^{c}_{\mu\lambda_1}\xi^{b},\nonumber\\
\delta_{\xi} A^{a}_{\mu\lambda_1 \lambda_2}& =&  ( \delta^{ab}\partial_{\mu}
+g f^{acb} A^{c}_{\mu})\xi^{b}_{\lambda_1\lambda_2} +
g f^{acb}(  A^{c}_{\mu  \lambda_1}\xi^{b}_{\lambda_2 } +
A^{c}_{\mu \lambda_2 }\xi^{b}_{\lambda_1}+
A^{c}_{\mu\lambda_1\lambda_2}\xi^{b}),\nn\\
.........&.&............................\nn
\eeqa
and the {\it complementary gauge transformation}   $\tilde{\delta}_{\eta}$
is \cite{Savvidy:2005ki} :
\beqa\label{doublepolygaugesymmetric}
\tilde{\delta}_{\eta} A^{a}_{\mu} &=& ( \delta^{ab}\partial_{\mu}
+g f^{acb}A^{c}_{\mu})\eta^b ,\\
\tilde{\delta}_{\eta} A^{a}_{\mu\lambda_1} &=&  ( \delta^{ab}\partial_{\lambda_1}
+  g f^{acb}A^{c}_{\lambda_1})\eta^{b}_{\mu} + g f^{acb}A^{c}_{\mu\lambda_1}\eta^{b},\nn\\
\tilde{\delta}_{\eta} A^{a}_{\mu\lambda_1\lambda_2} &=& ( \delta^{ab}\partial_{\lambda_1}
+g f^{acb} A^{c}_{\lambda_1})\eta^{b}_{\mu\lambda_2} +( \delta^{ab}\partial_{\lambda_2}
+g f^{acb} A^{c}_{\lambda_2})\eta^{b}_{\mu\lambda_1} +\nn\\
&~&
+g f^{acb}(  A^{c}_{\mu  \lambda_1}\eta^{b}_{\lambda_2 }+
A^{c}_{\mu \lambda_2 }\eta^{b}_{\lambda_1}+
A^{c}_{\lambda_1\lambda_2}\eta^{b}_{\mu} +
A^{c}_{\lambda_2 \lambda_1}\eta^{b}_{\mu} +A^{c}_{\mu\lambda_1\lambda_2}\eta^{b}),\nn\\
.........&.&............................\nn
\eeqa
These transformations form  a closed algebraic structure in the sense that
\beqa\label{gaugecommutator}
[~\delta_{\eta},\delta_{\xi}]~A_{\mu\lambda_1\lambda_2 ...\lambda_s} ~=~
-i g~ \delta_{\zeta} A_{\mu\lambda_1\lambda_2 ...\lambda_s},~~~~
[~\tilde{\delta}_{\eta},\tilde{\delta}_{\xi}]~A_{\mu\lambda_1\lambda_2 ...\lambda_s} ~=~
-i g~ \tilde{\delta}_{\zeta} A_{\mu\lambda_1\lambda_2 ...\lambda_s}\nn
\eeqa
and have the same composition law for the gauge parameters:
\beqa\label{gaugealgebra}
\zeta&=&[\eta,\xi]\\
\zeta_{\lambda_1}&=&[\eta,\xi_{\lambda_1}] +[\eta_{\lambda_1},\xi]\nn\\
\zeta_{\lambda_1\lambda_2} &=& [\eta,\xi_{\lambda_1\lambda_2}] +
[\eta_{\lambda_1},\xi_{\lambda_2}]
+ [\eta_{\lambda_2},\xi_{\lambda_1}]+[\eta_{\lambda_1\lambda_2},\xi],\nn\\
......&.&..........................\nn
\eeqa
The transformations $\delta_{\xi}$ and $\tilde{\delta}_{\eta}$ do not
coincide and are {\it complementary} to each other in the following sense:
in $\delta_{\xi}$
the derivatives of the gauge parameters $\{  \xi \}$ are  over the first
index $\mu$, while in $\tilde{\delta}_{\eta}$ the derivatives of the
gauge parameters $\{ \eta \}$ are over the rest of the totally symmetric
indices $\lambda_1 ... \lambda_{s}$, so that together they cover all
indices of the nonsymmetric tensor gauge fields $
A^{a}_{\mu\lambda_1 ... \lambda_{s}}$
(recall that these tensor gauge fields  are not symmetric with respect
to the index $\mu$ and the rest of the indices $\lambda_1 ... \lambda_{s}$).
Therefore the above transformations  (\ref{polygaugefull}) and (\ref{doublepolygaugesymmetric})
are complementary   representations of the same
infinite-dimensional gauge group ${\cal G}$ with the associative
algebra (\ref{gaugealgebra}) \cite{Savvidy:2005ki}.

The generalized field strength tensors $G^{a}_{\mu\nu,\lambda_1 ... \lambda_s}$
transform {\it homogeneously} with respect to the transformations $\delta_{\xi}$
(\ref{polygaugefull})
\cite{Savvidy:2005fi,Savvidy:2005zm}
and allow to construct the gauge invariant
Lagrangian ${\cal L}(A)$ which describes  dynamical tensor gauge bosons of all ranks
 \cite{Savvidy:2005fi,Savvidy:2005zm,Savvidy:2005ki,Savvidy:2005at}.
In recent publication \cite{Barrett:2007nn} the authors  constructed
complementary field strength tensors
$\tilde{G}^{a}_{\mu\nu,\lambda_1 ... \lambda_s}$
which are transforming  homogeneously,\footnote{The field strength
tensors $G^{a}_{\mu\nu,\lambda_1 ... \lambda_s}$ and
$\tilde{G}^{a}_{\mu\nu,\lambda_1 ... \lambda_s}$
are antisymmetric in their first two indices
and are totally symmetric with respect to the rest of the indices.
The explicit form of these tensors is given in
\cite{Savvidy:2005fi,Savvidy:2005zm,Barrett:2007nn}.}
now with respect to the $\tilde{\delta}_{\eta}$(\ref{doublepolygaugesymmetric})
\cite{Barrett:2007nn}
and  allow to construct the corresponding gauge invariant Lagrangian
$\tilde{{\cal L}}(A)$.

Thus there are two Lagrangian forms ${{\cal L}}(A)$  and
$\tilde{{\cal L}}(A)$ for the same tensor gauge fields $
A^{a}_{\mu\lambda_1 ... \lambda_{s}}$ which
are fully invariant with respect
to the corresponding gauge transformations (\ref{polygaugefull}) and
(\ref{doublepolygaugesymmetric}) \cite{Barrett:2007nn}
\beqa
\delta_{\xi}{{\cal L}}(A)=0, ~~~~~~~\tilde{\delta}_{\eta}\tilde{{\cal L}}(A)=0.\nn
\eeqa
The natural question which was raised at this point was to find out a
possible relation between these Lagrangian forms.
It has been found  that the following duality transformation\cite{Barrett:2007nn}
\beqa\label{dualtransformationincompo}
\begin{array}{ll}
\tilde{A}_{\mu\lambda_1} =  A_{\lambda_1\mu}   ,  \\
\tilde{A}_{\mu\lambda_1\lambda_2} =
{1\over 2}(A_{\lambda_1\mu\lambda_2} + A_{\lambda_2\mu\lambda_1})
-{1\over 2} A_{\mu\lambda_1\lambda_2}, \\
\tilde{A}_{\mu\lambda_1\lambda_2\lambda_3} =
{1\over 3}(A_{\lambda_1\mu\lambda_2\lambda_3}
+ A_{\lambda_2\mu\lambda_1\lambda_3}
+ A_{\lambda_3\mu\lambda_1\lambda_2})
-{2\over 3} A_{\mu\lambda_1\lambda_2 \lambda_3 }, \\
.........................................
 \end{array}
\eeqa
maps the Lagrangian $\tilde{{\cal L}}(A)$
into the Lagrangian ${{\cal L}}(\tilde{A})$. This takes place
because \cite{Barrett:2007nn}
$$
\tilde{G}_{\mu\nu,\lambda_1 ... \lambda_s}(A)  =
G_{\mu\nu,\lambda_1 ... \lambda_s}(\tilde{A})
$$
and therefore
$$
\tilde{{{\cal L}}}(A) ~~= ~~{{\cal L}}(\tilde{A}).
$$
One can find also  the inverse duality transformation  \cite{Barrett:2007nn}
\be\label{inversedualitymap}
 \begin{array}{ll}
A_{\mu\lambda_1} =  \tilde{A}_{\lambda_1\mu}   ,  \\
A_{\mu\lambda_1\lambda_2} =   \tilde{A}_{\lambda_1 \mu\lambda_2} +
\tilde{A}_{\lambda_2\mu\lambda_1},\\
A_{\mu\lambda_1\lambda_2\lambda_3} =
{1\over 3}(\tilde{A}_{\lambda_1\mu\lambda_2\lambda_3}
+ \tilde{A}_{\lambda_2\mu\lambda_1\lambda_3}
+ \tilde{A}_{\lambda_3\mu\lambda_1\lambda_2}),\\
..............
 \end{array}
\ee
The duality map (\ref{dualtransformationincompo}) is one-to-one.
The inverse transformation (\ref{inversedualitymap})
has the following unusual property.
If one applies the inverse transformation (\ref{inversedualitymap}) now
to the Lagrangian form $\CL(A)$, one can see that
the resulting expression can not be
identified with the Lagrangian form $\tilde{\CL}(\tilde{A})$, that is,
 $$\CL(A)~~\not\rightarrow~~\tilde{\CL}(\tilde{A}).$$

In this article we would like to find out an explanation for this phenomenon.
As we shall
demonstrate  there exists an infinite family (\ref{alternativedual})
of duality transformations
between tensor gauge fields. Within this family of duality transformations
there is a unique one  which has the
property that the direct and the inverse transformations coincide.
 This duality transformation
(\ref{alternative}),
(\ref{alternativedualtransformation}),
(\ref{dualityfieldtransformations})  has the desired property
that the direct and the inverse transformations map $\CL$ to $\tilde{\CL}$
and via versa
\beqa\label{alternativedualmap}
{{\cal L}}~~\leftrightarrow~~\tilde{{{\cal L}}} .\nn
\eeqa

\section{\it Two-Parameter Family of Duality Transformations}

The general form of the duality  transformation (\ref{dualtransformationincompo})
is \cite{Barrett:2007nn}
\be\label{dualtransformation}
\tilde{A}_{\mu\lambda_1 ...  \lambda_s} =
{1\over s}(A_{\lambda_1\mu...\lambda_s}
+ ....
+ A_{\lambda_s\mu... \lambda_{s-1}})
-{s-1\over s} A_{\mu\lambda_1 ... \lambda_s }~~~~s=1,2,.....
\ee
and can be expressed in the matrix form
\be\label{matrix}
\tilde{A}_{\mu\lambda_1...\lambda_s} = M_{\mu\lambda_1...\lambda_s}^
{~~~~~~~\nu\rho_1...\rho_s} ~A_{\nu\rho_1...\rho_s}~,
\ee
where the matrix $M$ and its inverse have the following structure:
\be\label{matrixformdual}
M= {1\over s} P - {s-1 \over s}\one ,~~~~~~ M^{-1}=P.
\ee
Here we have two operators, the permutation operator $P$ and the
identity operator $\one$. The permutation operator $P$ interchanges
the index $\mu$ with the indices  $\lambda_i$  and sums the result over all
$i=1,2,...,s$.
We see that the inverse transformation $M^{-1}$ does not coincide with the
direct transformation $M$. This is the main obstacle preventing the inverse
duality map to relate the Lagrangian form ${{\cal L}}$ with
$\tilde{{{\cal L}}}$, that is,
$\CL(A)~~\not\rightarrow~~\tilde{\CL}(\tilde{A})$.

Let us consider the properties of the   matrix $M$ in more details.
The permutation matrix $P$ has the
property
\be\label{square}
P^2 = (s-1) P + s\one
\ee
and therefore from (\ref{matrixformdual})
we can get the square of the duality
matrix $M$
\be
M^2 = {1-s \over s} M + {1 \over s}\one.
\ee
From this relation we can clearly see that the square of the duality matrix
$M$ is not equal to one: $M^2 \neq \one$. We would like to find out an
alternative duality
map $T$ for which $T^2 =\one$ and therefore $T^{-1}=T$.

The duality transformation (\ref{matrix}), (\ref{matrixformdual})
is a linear transformation of the basic tensor gauge fields and
can be defined by any nonsingular
matrix $T$:
\be\label{alternativedual}
\tilde{A}_{\mu\lambda_1...\lambda_s} = T_{\mu\lambda_1...\lambda_s}^
{~~~~~~~\nu\rho_1...\rho_s} ~A_{\nu\rho_1...\rho_s}.
\ee
Let us consider a two $(a,b)$-parameter class of  maps
similar to (\ref{matrixformdual})
\beqa\label{dualfamily}
T = a P + b\one. \nn
\eeqa
Calculating the square of the matrix $T$ and using the relation (\ref{square})
we  get
\beqa
T^2 = (2ab +a^2(s-1)) P + (b^2 + a^2 s)\one .\nn
\eeqa
Requiring that it is equal to the identity matrix we shall get a
system of algebraic equations:
\beqa
2ab +a^2(s-1)=0,~~~b^2 + a^2 s=1.\nn
\eeqa
The nontrivial solution gives the desired solution for $T$:
\be\label{alternative}
T  =  {2\over s +1} P - {s-1 \over s+1}\one.
\ee
This matrix has the property that $T^2 =1$ and therefore $T^{-1} =T$.
Thus the  transformation
(\ref{alternativedual}) will take the form
\be\label{alternativedualtransformation}
\tilde{A}_{\mu\lambda_1 ...  \lambda_s} =
{2\over s +1}(A_{\lambda_1\mu...\lambda_s}
+ ....
+ A_{\lambda_s\mu... \lambda_{s-1}})
-{s-1\over s+1} A_{\mu\lambda_1 ... \lambda_s }~~~~s=1,2,.....
\ee
In particular, for the first values of $s$, we have
\beqa\label{dualityfieldtransformations}
  \begin{array}{ll}
\tilde{A}_{\mu\lambda_1} =  A_{\lambda_1\mu}   ,  \\
\tilde{A}_{\mu\lambda_1\lambda_2} =
{2\over 3}(A_{\lambda_1\mu\lambda_2} + A_{\lambda_2\mu\lambda_1})
-{1\over 3} A_{\mu\lambda_1\lambda_2}, \\
\tilde{A}_{\mu\lambda_1\lambda_2\lambda_3} =
{1\over 2}(A_{\lambda_1\mu\lambda_2\lambda_3}
+ A_{\lambda_2\mu\lambda_1\lambda_3}
+ A_{\lambda_3\mu\lambda_1\lambda_2})
-{1\over 2} A_{\mu\lambda_1\lambda_2 \lambda_3 },  \\
.......................................
\end{array}
\eeqa
A posteriori one can get convinced that this duality map and its inverse
coincide. This is the main difference between duality maps
(\ref{dualtransformationincompo}),(\ref{dualtransformation})
and (\ref{alternativedualtransformation}),(\ref{dualityfieldtransformations}).

In (\ref{dualityfieldtransformations}) the first line defines the ordinary {\it transposition} and
the subsequent lines define natural generalization of the transposition operation
to the higher-dimensional tensors.
In the next section we shall demonstrate that there is an infinite family
of complementary gauge transformations
$\tilde{\delta}_\eta$ which have the same structure
as the  complementary gauge transformation (\ref{doublepolygaugesymmetric})
and that the above transformation  (\ref{alternative}),
(\ref{alternativedualtransformation}),
(\ref{dualityfieldtransformations}) defines a
natural duality map between them.

\section{\it Complementary Gauge Transformations}

The observation made in \cite{Barrett:2007nn}, that the complementary gauge
transformation (\ref{doublepolygaugesymmetric}) acting on the tensor gauge
field $\tilde{A}$ is identical with the
extended gauge transformation (\ref{polygaugefull}),
implies that the duality map serves
as a  similarity  transformation
between two representations of the same gauge algebra
(\ref{gaugealgebra}) (see also comment after formula (\ref{gaugealgebra})).
Therefore requiring that the gauge field  $\tilde{\CA}(e)$ \cite{Savvidy:2005ki}
transforms
by the extended gauge transformation  (\ref{polygaugefull}) we can find
out the complementary gauge transformations of the tensor gauge field $\CA(e)$
in the following  form\footnote{The symmetric indices
$\lambda_1,...,\lambda_s; \rho_1...\rho_s$ are suppressed in this formula
$T_{\mu\lambda_1...\lambda_s}^
{~~~~~~~\nu\rho_1...\rho_s} \sim T_{\mu }^
{~~ \nu }$\cite{Savvidy:2005ki}.}:
\be
\tilde{\delta}_{\eta}  \CA_{\mu} = T^{~\nu}_{\mu}~\partial_{\nu}\eta
-ig T^{~\rho}_{\mu} ~[T^{~\nu}_{\rho} \CA_{\nu} , \eta ].
\ee
To operate with this general formula one should expand the field $\CA(e)$
as in \cite{Savvidy:2005ki} and use the explicit form
of the matrices $T$. The  explicit form of these matrices can be obtained from
(\ref{alternative}), (\ref{alternativedualtransformation}),
(\ref{dualityfieldtransformations})
\beqa
T_{\mu\lambda_1}^{~~\nu\rho_1}&=&\delta^{\rho_1}_{\mu} \delta^{\nu}_{\lambda_1},\nn\\
T_{\mu\lambda_1 \lambda_2}^{~~~~\nu\rho_1 \rho_2}&=& {2\over 3}
(\delta^{\rho_1}_{\mu} \delta^{\nu}_{\lambda_1}\delta^{\rho_2}_{\lambda_2}+
\delta^{\rho_2}_{\mu} \delta^{\rho_1}_{\lambda_1}\delta^{\nu}_{\lambda_2})-
{1\over 3}\delta^{\nu}_{\mu} \delta^{\rho_1}_{\lambda_1}\delta^{\rho_2}_{\lambda_2},
\nn\\
T_{\mu\lambda_1 \lambda_2 \lambda_3}^{~~~~\nu\rho_1 \rho_2 \rho_3}&=&
{1\over 2}
(\delta^{\rho_1}_{\mu} \delta^{\nu}_{\lambda_1}\delta^{\rho_2}_{\lambda_2}\delta^{\rho_3}_{\lambda_3}+
\delta^{\rho_2}_{\mu} \delta^{\rho_1}_{\lambda_1}\delta^{\nu}_{\lambda_2}\delta^{\rho_3}_{\lambda_3}+
\delta^{\rho_3}_{\mu} \delta^{\rho_1}_{\lambda_1}\delta^{\rho_2}_{\lambda_2}
\delta^{\nu}_{\lambda_3})-
{1\over 3}\delta^{\nu}_{\mu} \delta^{\rho_1}_{\lambda_1}\delta^{\rho_2}_{\lambda_2}\delta^{\rho_3}_{\lambda_3},
\nn\\
...........&&............
\eeqa
Thus the complementary  gauge transformation
 $\tilde{\delta}_{\eta}$ of the tensor gauge fields is
\beqa\label{matrixform}
\tilde{\delta}_{\eta}  A_{\mu} &=& \partial_{\mu}\eta -i g[A_{\mu},\eta],\nonumber\\
\tilde{\delta}_{\eta}  A_{\mu\lambda_1} &=& \partial_{\lambda_1}\eta_{\mu} -i g[A_{\lambda_1},\eta_{\mu}]
-i g [A_{\mu\lambda_1},\eta],\nonumber\\
\tilde{\delta}_{\eta}  A_{\mu\lambda_1\lambda_2} &=&
{2 \over 3}(\partial_{\lambda_1}\eta_{\mu\lambda_2}
-i g[A_{\lambda_1},\eta_{\mu\lambda_2}] +\partial_{\lambda_2}\eta_{\mu\lambda_1}
-i g[A_{\lambda_2},\eta_{\mu\lambda_1}])-{1 \over 3}(\partial_{\mu}\eta_{\lambda_1\lambda_2}
-i g[A_{\mu},\eta_{\lambda_1\lambda_2}])\nn\\
&-&i g{2 \over 3}[A_{\mu\lambda_1},\eta_{\lambda_2}]
-i g {2 \over 3}[A_{\lambda_1\lambda_2},\eta_{\mu}]
+i g {1 \over 3}[A_{\lambda_1\mu},\eta_{\lambda_2}]\nn\\
&-&i g {2 \over 3}[A_{\mu\lambda_2},\eta_{\lambda_1}]-
i g{2 \over 3}[A_{\lambda_2\lambda_1},\eta_{\mu}]
+i g {1 \over 3}[A_{\lambda_2\mu},\eta_{\lambda_1}]
-i g [A_{\mu\lambda_1\lambda_2},\eta],\nn\\
.........&&...................................
\eeqa
where we have used the matrix notation  $A_{\mu\lambda_1...\lambda_s} =
A^{a}_{\mu\lambda_1...\lambda_s} L^a$
\cite{Savvidy:2005fi,Savvidy:2005zm,Savvidy:2005ki}. It is instructive to
compare this  complementary gauge transformation  with
(\ref{polygaugefull}) and (\ref{doublepolygaugesymmetric}).
At zero coupling constant, $g=0$, it gives
\beqa
\tilde{\delta}_{\eta}  A_{\mu} &=& \partial_{\mu}\eta ,\nonumber\\
\tilde{\delta}_{\eta}  A_{\mu\lambda_1} &=& \partial_{\lambda_1}\eta_{\mu} ,\nonumber\\
\tilde{\delta}_{\eta}  A_{\mu\lambda_1\lambda_2} &=&
{2 \over 3}(\partial_{\lambda_1}\eta_{\mu\lambda_2}
 +\partial_{\lambda_2}\eta_{\mu\lambda_1}
 )-{1 \over 3}\partial_{\mu}\eta_{\lambda_1\lambda_2},\nn\\
.........&&...................................
\eeqa
and  defines the behavior of the longitudinal parts of the
tensor gauge field with respect to the symmetric indices
$\lambda_1,...\lambda_s$,
in a  way similar to (\ref{polygaugefull}) and (\ref{doublepolygaugesymmetric}).
The corresponding covariant field strength tensors can also be constructed
with the use of the matrix $T$:
\beqa\label{fieldstrengtselfduial}
\tilde{G}_{\mu\nu} &\equiv& G_{\mu\nu}=
\partial_{\mu} A_{\nu} - \partial_{\nu} A_{\mu} -
ig  [~A_{\mu}~A_{\nu}],\\
\tilde{G}_{\mu\nu,\lambda_1} &=&
\partial_{\mu} A_{\lambda_1\nu} - \partial_{\nu} A_{\lambda_1\mu} -
i g  [~A_{\mu}~A_{\lambda_1\nu}] -ig [~ A_{\lambda_1\mu}~A_{\nu}],
\nn\\
\tilde{G}_{\mu\nu,\lambda_1\lambda_2} &=&  ~
\partial_{\mu}( {2\over 3}A_{\lambda_1\nu\lambda_2}+
{2\over 3}A_{\lambda_2\nu\lambda_1} - {1\over 3}A_{\nu\lambda_1\lambda_2})
-i g  [~A_{\mu},~
( {2\over 3}A_{\lambda_1\nu\lambda_2}+
{2\over 3}A_{\lambda_2\nu\lambda_1} - {1\over 3}A_{\nu\lambda_1\lambda_2})] -\nn\\
&-& \partial_{\nu} ( {2\over 3}A_{\lambda_1\mu\lambda_2}
+ {2\over 3}A_{\lambda_2\mu\lambda_1}- {1\over 3} A_{\mu\lambda_1\lambda_2})
-ig~[(  {2\over 3}A_{\lambda_1\mu\lambda_2}
+ {2\over 3}A_{\lambda_2\mu\lambda_1}-
{1\over 3} A_{\mu\lambda_1\lambda_2}),~A_{\nu}]
\nn\\
&-& ig  ~[~A_{\lambda_1\mu},~A_{\lambda_2\nu}]
-ig[~A_{\lambda_2\mu},~A_{\lambda_1\nu}],\nn\\
.......&&.........................\nn
\eeqa
These field strength tensors transform homogeneously and allow to construct
the invariant Lagrangian $\tilde{\CL}(A)$ quadratic in field strength
tensors
\cite{Savvidy:2005fi,Savvidy:2005zm,Savvidy:2005ki,Barrett:2007nn}.
The dual transformation  (\ref{alternativedualtransformation}),
(\ref{dualityfieldtransformations}) tells  us  now that
\beqa
\tilde{G}_{\mu\nu,\lambda_1 ... \lambda_s}(A)  =
G_{\mu\nu,\lambda_1 ... \lambda_s}(\tilde{A})
\eeqa
and therefore
\be
\tilde{\CL}(A) = \CL(\tilde{A}).
\ee
But now, with the  duality transformation
(\ref{alternativedualtransformation}), we shall have the
{\it additional property} under the inverse transformation $T^{-1}=T$
 (\ref{alternativedualtransformation})
\beqa
G_{\mu\nu,\lambda_1 ... \lambda_s}(A)  =
\tilde{G}_{\mu\nu,\lambda_1 ... \lambda_s}(\tilde{A})
\eeqa
which is easy to check using (\ref{fieldstrengtselfduial}) and the
definition of $G_{\mu\nu,\lambda_1 ... \lambda_s}(A)$
\cite{Savvidy:2005fi,Savvidy:2005zm,Savvidy:2005ki}.
Therefore we have now the desired property that
\be
\CL(A) = \tilde{\CL}(\tilde{A}).
\ee
This solves the problem posed in the introduction. In the next
two sections we shall discuss some additional properties of this
duality map and the complementary gauge transformations.

\section{\it Duality Transformation of High Forms}

As we have seen above the duality transformation
(\ref{alternative}),
(\ref{alternativedualtransformation}),
(\ref{dualityfieldtransformations}) for the rank-2  tensors (s=1)
coincides
with the {\it ordinary transposition} of the matrices, which has the
property that it squares to  $\one$.
Our duality operator $T$,  defined above  (\ref{alternative}),
(\ref{alternativedualtransformation}),
(\ref{dualityfieldtransformations})
also has the same property $T^2 =\one$ and can be considered therefore
as a natural generalization
of the transposition operation to the higher-dimensional tensors.
This observation allows to define symmetric and antisymmetric tensor
gauge fields. Indeed
as for the ordinary transposition, we can use the generalized transposition,
to define symmetric and  antisymmetric  parts of a higher-rank
tensor gauge field  $A$ as
$$
A^{\rm sym} = \frac12(A+\tilde{A} ),\qquad A^{\rm asym}=
\frac12(A-\tilde{A} )\quad.
$$
They are symmetric or antisymmetric with respect to the generalized transposition.
Their explicit forms are
\beqa
(A^{\rm sym})_{\mu\lambda_1 ...  \lambda_s} &=&
{1\over s +1}(A_{\mu\lambda_1 ...
\lambda_s }+A_{\lambda_1\mu\lambda_2...\lambda_s}
+ ....
+ A_{\lambda_s\lambda_1\ldots\lambda_{s-1}\mu})\nn\\
(A^{\rm asym})_{\mu\lambda_1 ...  \lambda_s} &=&
{s\over s+1} A_{\mu\lambda_1 ... \lambda_s }-
{1\over s +1}(A_{\lambda_1\mu\lambda_2...\lambda_s}
+ ....
+ A_{\lambda_s\lambda_1\ldots\lambda_{s-1}\mu})
\nn\eeqa
The symmetric part $A^{\rm sym}$ thus indeed coincides with the total
symmetrization of all indices. The total antisymmetrization would of
course vanish, as the gauge field is symmetric in all but one index,
but $A^{\rm asym}$ is as antisymmetric as it can be.

It is an interesting question if one can generalize  this transposition
operation to higher-degree forms. The
field strength tensor  $G_{\mu\nu,\lambda_1\ldots\lambda_s}$
can be an important example. Let us consider a tensor
$G_{\mu_1\ldots\mu_n,\lambda_1...\lambda_s}$ which
is antisymmetric in its first $n$ indices and is symmetric in the
following $s$ indices.
The sum $n+s$ is the tensor rank. A natural generalization of the index
permutation $P$ is given by
\be
(P G)_{\mu_1\ldots\mu_n,\lambda_1\ldots\lambda_s}
= \sum_{i=1}^n\sum_{j=1}^s
G_{\mu_1\ldots\mu_{i-1}\lambda_j\mu_{i+1}\ldots\mu_{n},\lambda_1
\ldots\lambda_{j-1}\mu_i\lambda_{j+1}\ldots\lambda_s}.
\ee
Calculating its square one can get
\be
P^2=(s-n)P+sn\one \quad.
\ee
Again considering a two-parameter family of operators
$$
T = a P + b \one
$$
one can get a unique
linear combination of $P$ and $\one$ which squares to $\one$.
It has the form
\be
T=  \frac {2}{s+n}P-\frac{s-n}{s+n}\one \quad.
\ee
For the field strength tensors of the lower rank we shall have
\beqa
G^{T}_{\mu \nu,\lambda_1 }&=& {2\over 3}G_{\mu\lambda_1,\nu}
+{2\over 3}G_{\lambda_1\nu,\mu} +{1\over 3}G_{\mu\nu,\lambda_1} \nn\\
G^{T}_{\mu \nu,\lambda_1\lambda_2 }&=& {1\over 2}G_{\mu\lambda_1,\nu\lambda_2}
+{1\over 2}G_{\lambda_1\nu,\mu\lambda_2} +{1\over 2}G_{\mu\lambda_2,\lambda_1\nu}
 +{1\over 2}G_{\lambda_2\nu,\lambda_1\mu} \nn\\
 ........&&.............................
\eeqa
The above  transposition law can be used
to define symmetric and antisymmetric parts of the
field strength tensors, but it is not clear yet,  what is
the role of the above construction in  the generalization
of gauge field theory.

\section{\it Enhanced Gauge Algebra}

It was observed in \cite{Savvidy:2005fi,Savvidy:2005ki} that for a
certain linear combination of Lagrangian forms the rank two
gauge field  exhibits an enhanced gauge symmetry,
where the extended and the complementary gauge transformations are
realized at the same time.
It is therefore interesting to check, whether the sum of extended and of the
complementary gauge transformations forms  a closed algebra.
The sum has the form\footnote{The definition is $\nabla_{\mu}\xi_{\lambda }=
\partial_\mu \xi_{\lambda} -i g [A_\mu , \xi_{\lambda}]$.}
\be
(\delta_{\xi}+\tilde{\delta}_{\eta})A_{\mu\lambda_1}=
 \nabla_{\mu}\xi_{\lambda_1} +
 \nabla_{\lambda_1}\eta_{\mu}
-ig [A_{\mu\lambda_1}, \xi + \eta ]
\ee
and the commutator of two such sums is
$$
[\delta_{\xi}+\tilde{\delta}_{\eta}, \delta_{\psi}+\tilde{\delta}_{\chi}]
A_{\mu\lambda_1}
$$
and contains, in particular, the commutator of
the extended gauge transformation with the complementary one which we can
easily compute
\be
[\tilde{\delta}_{\eta},\delta_{\xi}] A_{\mu\lambda_1}= -ig
\{\nabla_{\mu}[\eta , \xi_{\lambda_1}] + \nabla_{\lambda_1}[\eta_\mu , \xi]
-ig [A_{\mu\lambda_1}, [\eta, \xi]\}.
\ee
Now we can check that the algebra is closed
\beqa
[\delta_{\xi}+\tilde{\delta}_{\eta}, \delta_{\psi}+\tilde{\delta}_{\chi}]
A_{\mu\lambda_1}=([\delta_{\xi},\delta_{\psi}] +
[\tilde{\delta}_{\eta},\tilde{\delta}_{\chi}] +
[\delta_{\xi},\tilde{\delta}_{\chi}]+
[\tilde{\delta}_{\eta},\delta_{\psi}])
A_{\mu\lambda_1}=\nn\\
-ig
\{\nabla_{\mu}([\xi, \psi_{\lambda_1}] + [\xi_{\lambda_1}, \psi ]+
[\eta, \psi_{\lambda_1}] + [\xi_{\lambda_1}, \chi ]) +\nn\\
+ \nabla_{\lambda_1}([\eta, \chi_{\mu}] + [\eta_{\mu}, \chi ]+
[\eta_{\mu}, \psi] + [\xi, \chi_{\mu} ])-\nn\\
-ig [A_{\mu\lambda_1}, ([\xi,\psi ]+[\eta, \chi]+[\eta, \psi]+[\xi, \chi])\}\nn\\
=-ig\{\nabla_{\mu}\zeta_{\lambda_1} +
 \nabla_{\lambda_1}\omega_{\mu}
-ig [A_{\mu\lambda_1}, \varphi ]\}.
\eeqa
Thus indeed it is a similar transformation with the gauge parameters
\beqa
\zeta_{\lambda_1}=[\xi, \psi_{\lambda_1}] + [\xi_{\lambda_1}, \psi ]+
[\eta, \psi_{\lambda_1}] + [\xi_{\lambda_1}, \chi ]=
[\xi + \eta, \psi_{\lambda_1}] + [\xi_{\lambda_1}, \psi +\chi]\nn\\
\omega_{\mu}=[\eta, \chi_{\mu}] + [\eta_{\mu}, \chi ]+
[\eta_{\mu}, \psi] + [\xi, \chi_{\mu} ]=
[\xi + \eta, \chi_{\mu}] + [\eta_{\mu},\psi + \chi]\nn\\
\varphi = [\xi,\psi ]+[\eta, \chi]+[\eta, \psi]+[\xi, \chi]=
[\xi + \eta, \psi +\chi]
\eeqa

\section{\it Acknowledgement}
The work of (S.G.) was supported by ENRAGE (European Network on Random
Geometry), a Marie Curie Research Training Network, contract MRTN-CT-2004-
005616. The work of (G.S.) was partially supported by the EEC Grant
no. MRTN-CT-2004-005616.

\end{document}